\documentclass[runningheads]{llncs}
\usepackage[T1]{fontenc}
\usepackage{graphicx}
\usepackage{amsmath,amssymb,amsfonts}
\usepackage{cite}
\usepackage{textcomp}
\usepackage{xcolor}
\usepackage{listings}
\usepackage{xcolor}
\usepackage{float}
\usepackage{xspace}
\usepackage{listings}
\usepackage{verbatim}
\usepackage{verbatimbox}
\usepackage{mdframed}
\usepackage{array}
\usepackage{dirtree}
\usepackage{algorithm}
\usepackage{algpseudocode}
\usepackage{graphicx}
\usepackage[font=small,skip=0pt]{caption}
\usepackage{subcaption}
\usepackage{booktabs}
\usepackage{xurl}
\usepackage[shortlabels]{enumitem}
\usepackage{balance}
\usepackage{multicol}
\usepackage[most]{tcolorbox}
\newcommand{\rqbox}[1]{
\begin{tcolorbox}[tile, size=fbox, boxsep=2mm, boxrule=0pt, top=0pt, bottom=0pt,
borderline west={1mm}{0pt}{blue!50!white}, colback=blue!5!white]
#1
\end{tcolorbox}
}

\newcommand{\ourapp}{{RustMap}\xspace}

\usepackage{soul}  
\usepackage[disable,colorinlistoftodos,prependcaption,textsize=tiny]{todonotes} 
\usepackage{marginnote}

\newcommand{\edits}[1]{#1}
\newcommand{\basicalert}[2]{\fbox{\bfseries\sffamily\scriptsize\color{blue} #1}{\sf\small$\blacktriangleright$\textit{\color{red} #2}$\blacktriangleleft$}}
\newcommand{\jkalert}[2]{\fbox{\bfseries\sffamily\scriptsize\color{red} #1}{\sf\small$\blacktriangleright$\textit{\color{blue} #2}$\blacktriangleleft$}}
\newcommand{\jk}[1]{\jkalert{From Jiakun}{#1}}
\newcommand{\lx}[1]{\basicalert{From JLX}{#1}}

\newcommand{\xm}[1]{\basicalert{From CXM}{#1}}

\begin{document}
\title{RustMap: Towards Project-Scale C-to-Rust Migration via Program Analysis and LLM}
\titlerunning{RustMap: Project-Scale C-to-Rust Migration}
%
\author{Xuemeng Cai\inst{1} \and Jiakun Liu\inst{1} \and Xiping Huang\inst{1} \and Yijun Yu\inst{2} \and Haitao Wu\inst{3} \and Chunmiao Li\inst{4} \and Bo Wang\inst{5} \and Imam Nur Bani Yusuf\inst{1} \and Lingxiao Jiang\inst{1}}
%
\authorrunning{Cai et al.}
%
\institute{Singapore Management University \email{\{xuemengcai,jkliu,xphuang,imamy.2020,lxjiang\}@smu.edu.sg} \and
The Open University \email{y.yu@open.ac.uk} \and 
Independent Researcher \email{wain303009@hotmail.com} \and 
Beijing Academy of Blockchain and Edge Computing \email{chunmiaoli1993@gmail.com} \and 
Beijing Jiaotong University \email{wangbo\_cs@bjtu.edu.cn}}
%
\maketitle              
\begin{abstract}
Migrating existing C programs into Rust is increasingly desired, as Rust offers superior memory safety while maintaining C's high performance.
However, vastly different features between C and Rust, e.g., distinct definitions and usages of pointers and references, pose significant challenges beyond mere syntactic translation.
Existing automated translation tools, such as C2Rust, may rely too much on syntactic, template-based translation and generate unsafe Rust code that is hard for human developers to read, maintain, or even compile.
More semantic-aware translation that produces safer, idiomatic, and runnable Rust code is much needed.
This paper introduces a novel dependency-guided and large language model (LLM)-based C-to-Rust translation approach, \ourapp, based on three key ideas:
(1) Utilize LLM's capabilities to produce idiomatic Rust code from given \emph{small} pieces of C code,
(2) Mitigate LLM's incapability in handling large codebases by breaking project-scale C programs into smaller units for translation according to their usage dependencies,
and later composing them together into a runnable Rust program,
and (3) Enhance the correctness of the translated Rust program by utilizing available test cases to check input/output equivalence between C and Rust code, 
isolating faulty code when program execution states deviate, 
and iteratively utilizing the feedback from compilation and testing errors for LLM to refine translated Rust code.
We have empirically evaluated \ourapp on 126 sample real-world programs,
including 125 programs from Rosetta Code and a complex bzip2 program containing more than 7000 lines of code, using GPT-4o as the LLM. \ourapp shows promising results in guiding GPT-4o to translate most of the C code into more idiomatic, readable, and functional Rust code with significantly less unsafe code than other translation tools,
presenting non-trivial translation patterns that may be reusable for future research.
Our study also identifies typical cases that are difficult for other tools when translating C code into Rust 
and presents possible solutions constructed by \ourapp that may be reusable for future research.

\end{abstract}

\section{Introduction}
\label{sec:intro}

Rust is increasingly popular as it not only offers better safety guarantees
but also has performance on par with that of C~\cite{bang2023trust, memorysafetyrust2020}.
In industry, there is a growing trend to rewrite existing C projects in Rust \cite{TaKO8Ki2024, casey2024, infoq2023rusttrend}, including the Linux kernel \cite{rust4linux, redox}.
To help developers to translate C code into Rust, several tools have been proposed.
For example, C2Rust \cite{c2rust_challenges} aims to automatically convert C code to Rust with mostly syntax-based rules.
However, such tools often produce too much unsafe code that is not idiomatic in Rust and hard to read or maintain since C and Rust languages have very different features, e.g., object ownership and lifetime rules.
To tackle the drawbacks, more tools, such as CRustS~\cite{crusts}, Crown~\cite{crown}, and Laertes~\cite{laertes21, laertes23},  have been proposed on top of C2Rust to analyze specific C \& Rust language features and enhance transformation between them.
However, due to the lack of one-to-one syntactic mapping between C and Rust language features (e.g., incompatible data types, macros, vastly different usages of pointers and references, language-specific programming idioms, and Rust-specific ownership and lifetime rules), 
translation between them is not trivial and these tools are still very limited in translating diverse code patterns, and the translated Rust code still contain lots of unsafe code blocks that are hard to read, maintain, or even compile---Challenge \textcircled{1}.
In addition, these tools often fail to translate project-scale C programs to produce executable Rust programs, due to inadequate management of various kinds of dependencies among code, such as language-specific libraries, linker files, global data types and variables defined in many header files scattered across different locations in the project but dependent on by a piece of code to be translated. When a C project is translated to Rust, the resultant Rust project structure should accommodate all the dependent code from C to ensure their compilation and execution---Challenge \textcircled{2}.
Also, the translated Rust code should naturally be functionally equivalent to the original C code---Challenge \textcircled{3},\footnote{except for the cases when the C code contains bugs.} 
i.e., producing the same outputs when executed with the same input.

To address these challenges, we propose a novel LLM-prompting-based translation approach guided by program dependency analyses, named {\bf \ourapp}.
We utilize the capabilities of a LLM (specifically ChatGPT \cite{white2023chatgpt,DBLP:journals/corr/abs-2309-08221}) for code translation \cite{yang2024exploring}
since LLMs has recently shown great potentials in software engineering tasks, including code-to-code translation \cite{yang2024exploring}, code-to-text summarization \cite{nam2024using,ahmed2024automatic,chen2021my}, and text-to-code search and generation \cite{bui2021infercode,xu2022systematic,gu2023llm,du2024mercury}, although LLMs are not yet able to handle large input and the code produced by LLMs can be buggy \cite{pan2024lost,liu2024your}.
To deal with LLM's limits in handling large input data and avoiding bugs \cite{pan2024lost,liu2024your},
\ourapp utilizes program dependency analyses and functional testing, combined with customized prompts, to extend LLM's translation capabilities to project-scale programs.

Specifically, \ourapp first breaks down a C program into relatively smaller and standalone translation units, 
which can be a global variable definition, a datatype \lstinline{struct} definition, a macro, a preprocessor directive, a function, or a set of inter-dependent functions,
by utilizing both static and dynamic analyses of the C source code and a given set of functional test cases.
Then \ourapp determines the translation order of these units
according to their dependencies,
based on the function execution sequences during tests, with additional considerations from the syntactic structures and control flows in source code.
It constructs a Rust project that contains sufficient Rust files or modules to hold all the code translated from the translation units for the whole-program translation.
This dependency-guided break-down of C programs and construction of Rust project structures, which we call \emph{project scaffolding} tackles the Challenges \textcircled{1} and \textcircled{2}:
(1) Suitable prompts for each translation unit are formulated and fed to ChatGPT for translation, utilizing ChatGPT's capabilities in producing functionally the same, but syntactically different code, so that the translated Rust code can be safer and more idiomatic. 
(2) With the broken-down smaller and relatively standalone translation units, \ourapp makes it possible to apply LLMs to large programs.
For Challenge \textcircled{3}, \ourapp also utilizes the functional tests together with ChatGPT to enhance the correctness of translated Rust code: Whenever there are compilation or testing errors in the translated Rust, the error information, either compilation error messages or code execution states, is extracted and formulated as suitable prompts for ChatGPT to locate and adjust faulty translated code until the code passes the provided functional tests or time out.
Individual translated units in Rust are composed together according to the dependencies of their counterparts in C to complete the whole-program migration.
Thus, \ourapp provides an operational roadmap for translating project-scale C programs into Rust through a LLM-prompting-based, dependency-guided translation process to enhance the correctness, readability, safety, executability of translated Rust programs.

We have evaluated \ourapp on 126 real-world C programs ranging from tens of lines to more than 7000 lines of code, including sample Rosetta Code \cite{rosettacode}, 
and a file compression program {\tt bzip2} \cite{bzip2}.
Our results show that \ourapp can translate most of the C programs into idiomatic Rust programs without the need of defining specific transformation rules.
Most programs translated by \ourapp are compilable and runnable, passing a set of functional tests, indicating the same functionalities as the original C programs. 
\ourapp also significantly reduces unsafe code and improves readability in the translated programs when compared with other translation tools.

Our study also empirically identifies a number of typical cases that fail to be translated by other tools (e.g., C2Rust), and provides solutions constructed by \ourapp.
For example, \ourapp identifies C pointers that can often be rewritten as vectors in Rust, pointer aliasing that can be replaced by memory copies, global variables that need to be wrapped into safer \lstinline{lazy_static} blocks, and C-style \lstinline{switch-cases} with fall-through that need to be converted into Rust-style while loops with \lstinline|match| statements.

In short, the main contribution of our paper is \ourapp, a novel dependency-guided, LLM-prompting-based approach for project-scale C to Rust translation. It leverages functional testing and dependency analyses to guide the project-scale translation process, utilizes LLMs to perform actual translation, produces more idiomatic and safer code than other tools, and enhance the correctness of translated code.
Our evaluation on small to medium real-world C programs of various complexities shows the advantages of our approach against other translation tools.
Although not without limitations,
our approach and the translated Rust code
for typical difficult cases faced by other tools provide meaningful and reusable solutions for future research and tools.
Our code and translation results are publicly available at \url{https://github.com/Cxm211/RustMap}.

\textbf{Paper organization.}
The remainder of this paper is organized as follows:
Section \ref{sec:overview} provides an overview of \ourapp.
Section~\ref{sec:scaffolding} introduces the project-level organization of translated code.
Section~\ref{sec:translation} describes more details about translating C code into Rust using LLM.
Section~\ref{sec:evaluation} evaluates our approach.
It also presents typical cases that are challenging for automatic translation,
and discusses our limitations and threats to validity. 
Section~\ref{sec:related} reviews related work.
Section \ref{sec:conclusion} concludes with future work.



\section{\ourapp Overview}
\label{sec:overview}

\begin{figure}[t]
    \centering
    \includegraphics[width=0.9\linewidth]{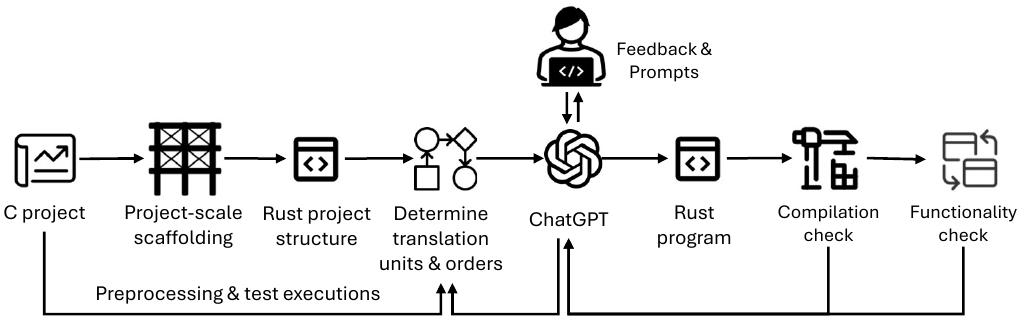} 
    \caption{Overview of \ourapp}
    \label{fig:rustmap_overview_diagram}
\end{figure}

Figure \ref{fig:rustmap_overview_diagram} illustrates an overview of \ourapp.
We focus on migrating a C project to Rust by translating C functions and their dependencies one by one.
Given a C project, \ourapp first generates a Rust project scaffolding where each C header file and each C function is mapped to a Rust file; the Rust files are initially empty, but when more C functions or header file contents are translated, the translated code will gradually be merged into the files.
\ourapp also preprocesses the C files, compiles the whole program, and run its available tests, which provides static and dynamic call graphs to identify the dependencies of each function (e.g., other functions, data types, global variables, and macros used by it). Sections \ref{sec:structuring} and \ref{sec:preprocessing} have more details.

Then, \ourapp determines the units for translation and their translation order based on 
dependencies among code elements. Each translation unit may have one or many code elements, such as {\tt struct} definitions, global variables, macros, a function definition, a statement, etc. The translation units and order can be adjusted according to LLM's outputs (e.g., when ChatGPT identifies more dependent code elements to be translated). Section \ref{sec:dependency-order} has more details.

A translation unit (e.g., a function and its dependencies) along with custom prompts is fed to LLM for translation (see Section \ref{sec:chatgpt-translation}).
If LLM fails to produce translated Rust code, we try again with another translation unit or adjust the translation unit (e.g., splitting a large unit into smaller ones) according to its outputs.
%
When the LLM produces Rust code for the unit, \ourapp applies the Rust compiler to check its syntactic correctness.
If there were any compilation errors
in the translated code, 
\ourapp will try to fix them by using the error messages as additional prompts to LLM and instructing ChatGPT to retranslate the code and fix the given error (see Section \ref{sec:compilation-errors}).

When all the units needed to run a given test have been translated, \ourapp will try to compile them together to produce an executable Rust program.
Similar to the above, if there were any compilation errors in the translated code, \ourapp will try to fix them by using the error messages as additional prompts to LLM.
\ourapp will execute the compiled Rust program with the given test input and compare its execution states with those of the original C program executed with the same test input to identify any semantic differences.
If there was any difference, \ourapp will try to locate the error and fix it by using the execution state context as additional prompts to LLM and instructing ChatGPT to retranslate the code so that the execution states in the translated Rust code can match their C counterparts (see Section \ref{sec:error-resolution}).

As a result, \ourapp will generate a project-scale Rust program with
sufficient C functions and their dependencies translated so that it can run the same given tests to produce the same outputs, while maintaining close mapping between the original C functions and the translated Rust code.

\section{Project-Scale Scaffolding}
\label{sec:scaffolding}
A project can contain a large amount of code, and it is impractical to translate the code all at once. When translating a C project to a Rust project, we temporarily organize the C project in a way that allows each C function and/or its relevant dependencies to be translated separately.
For this purpose, we need to identify the boundaries of \emph{translation units} where each unit is a set of C code and/or their dependencies to be translated together, and determine the order in which they can be translated.
We also construct a Rust project containing temporary files corresponding to the translation units, to save intermediate translation results, track dependencies, and prepare for integrated compilation and execution of the whole translated Rust program. 
We refer to such preparations as \textit{scaffolding}.
The folders and files generated during scaffolding are temporary, mainly for managing the translation units, and 
will be merged together so that the final Rust project structure closely matches the original C project file/folder structure.
The following subsections detail the preparations.

\subsection{Initial Rust Project Structure}
\label{sec:structuring}

Our overarching idea is to use a function and/or its dependencies as a translation unit and translate one or multiple units at a time as long as the LLM can handle them.
We use functions as translation units mainly because C programs start executions from their {\tt main} functions and functions are often natural modular units of code and have well-defined interfaces and dependencies.\footnote{For functions longer than the input length limit of LLM, we divide function bodies and their dependencies into smaller
but syntactically complete 
statements for translation.}
By translating functions individually, it is easier to trace and resolve their dependencies, and easier to integrate translated functions into the Rust project following the C project structure.

We use the \lstinline{cargo new} command to initialize a Rust project. The command generates a \lstinline{Cargo.toml} file, which contains the metadata about the project and its dependencies.
The metadata will be updated when new Rust files are generated during the translation and is used to compile and run the project.
After doing so, the project is initialized with the necessary files and directories, meanwhile, can be built and run.
For each \lstinline{.c} file in the given C project, we automatically map the C file to a set of Rust \lstinline{.rs} files.
Each Rust file corresponds to either a function from the C file, 
or a set of global variables, data types, or preprocessor directives and macros
\footnote{\edits{Note that we handle some of these code elements differently from others for better code safety and readability (see Sections \ref{sec:macro-refactoring} and \ref{sec:lazy-static}).}}
defined in the C file, and is put in a directory of the same name (without the \lstinline{.c} suffix) as the C file.
For each \lstinline{.h} header file in the C project, a corresponding \lstinline{.rs} file is also generated and placed in the \textit{headers} folder to be shared across the project.
If a \lstinline{.h} file also happens to include function definitions (in addition to function declarations), those functions are processed in the same way as \lstinline{.c} files.

\begin{figure}[t]
    \centering
    \includegraphics[width=0.95\linewidth]{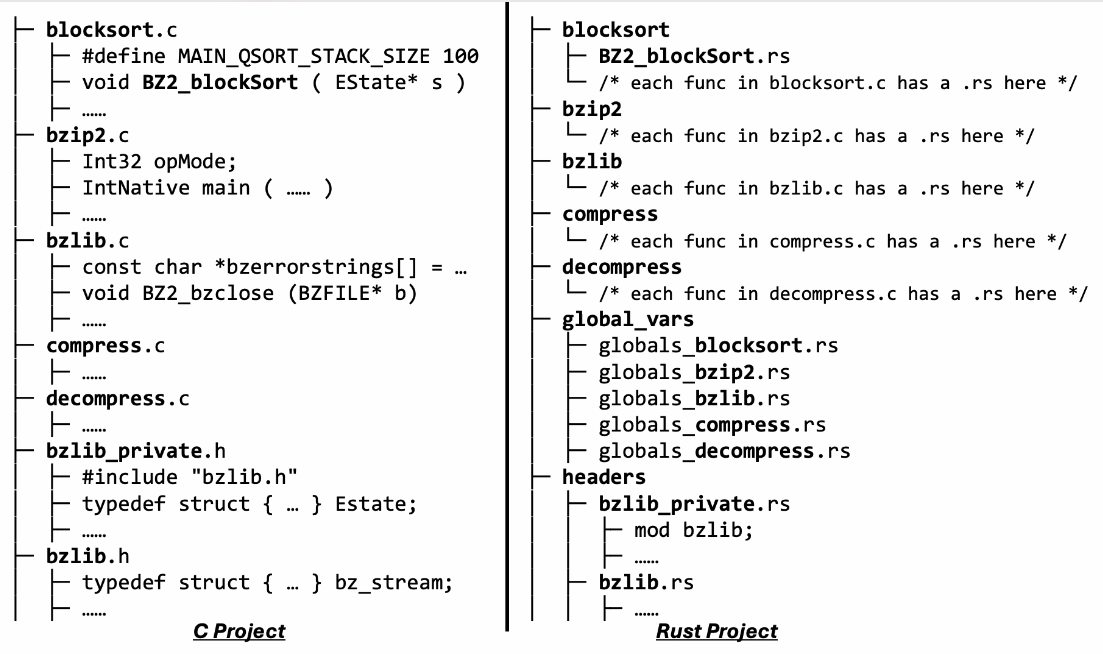} 
    \caption{An example of project scaffolding structure}
    \label{fig:headers_globals_synthesis}
\end{figure}

Figure \ref{fig:headers_globals_synthesis} is a sample generated Rust project scaffolding for the bzip2 program used in our evaluation,
where various scaffolding folders and files are created for storing code according to either globals, functions, or headers that will be handled as translation units.
The scaffolding files can be merged or split during translation, depending on how well LLM handles the translation units.
For example, the global variable definitions in \lstinline{blocksort.c} file will be translated and put into the \lstinline{global_vars/globals_blocksort.rs} file in Rust; the \lstinline{BZ2_blockSort} function in the C file will be translated and put into the \lstinline{blocksort/BZ2_blockSort.rs} file; etc.
For another example, the \lstinline{bzlib_private.h} file will be translated and put into the \lstinline{headers/bzip2_private.rs} file in Rust; since it includes \lstinline{bzlib.h} in C, the translated \lstinline{bzlib_private.rs} file will include \lstinline{mod bzlib;} so that it can use code elements defined in \lstinline{headers/bzlib.rs}.

\subsection{Handling C Preprocessor Directives}
\label{sec:preprocessing}

A C file can include many kinds of preprocessor directives, e.g., \lstinline|#include| for file inclusions, \lstinline|#define| for macro definition, \lstinline|#if| for conditional compilation, \lstinline|#pragma| for specific compilation features, etc., which may bring many challenges when being translated into Rust.
To the best of our knowledge, all prior C-to-Rust translation tools use C's preprocessor to turn a \lstinline|.c| file into a \lstinline|.i| file so that all directives are preprocessed and removed so that they do not actually translate directives.
However, preprocessed directives, especially macro expansions, can lead to much bloated code from the included libraries and much repetitive code that is less readable. It would be beneficial for users if the directives and macros can be translated to corresponding Rust code elements as much as possible to retain code readability and maintainability.

Thus, for each \lstinline|.c| file, we treat all the preprocessor directives as a translation unit, together with other code elements from the file that may directly depend on them according to a simple def-use static analysis, and feed them into LLM.
The unit may be split according to the def-use dependency order\footnote{Intuitively, we handle the declarations or definitions of directives first before handling other code elements that \emph{use} the definitions.} if it is too large to fit in the length limit of LLM.


\subsubsection{Handling file inclusions}
When the directives include commonly used libraries, such as \lstinline|#include <stdlib.h>|, we observe that ChatGPT can recognize the functions and macros defined in the libraries and used in the C code to be translated, and replace those C functions and macros with appropriate Rust library functions and macros. Thus, we can leave out such directives from further processing. 

For directives that include project-specific header files, e.g., \lstinline|#include "bzlib.h"|, we will translate each header file and replace the directive with a \lstinline|mod| declaration (translated as \lstinline|mod bzlib;|)
in the corresponding scaffolding file and later add a \lstinline|use| statement if 
an element from the header file is used somewhere (see Section \ref{sec:add-dependencies}), to indicate the dependencies.

\subsubsection{Handling C Macros}
\label{sec:macro-refactoring}

The existing tool C2Rust \cite{c2rustfaq} preprocesses C programs before translating C files to Rust. Thus, C2Rust expands all macro uses into concrete C code, including primitive-type numerical macros, complex macros \cite{arm_complex_macros}, and others, leading to duplicate and less readable code.
For better code readability and alignment with the original C code, we propose to retain the macro definitions as much as possible based on LLM's capabilities in translating some of C macros to Rust macros and simple rule-based refactoring.

%


\paragraph{Replacing numerical macros with const global variables.}

For primitive-type numerical macros, we use LLM to infer the data type of each numerical value and use rule-based automation to rewrite them into const global variables in Rust. A detailed example of this transformation can be found in the corresponding section\footnote{\label{example}\url{https://github.com/Cxm211/RustMap/blob/main/README_example.md}} in our Github repository.
Figure \ref{fig:numerical_macros_definition} shows sample numerical macros and their refactoring and translations. For example, \lstinline{#define BZ_RUNB 1} is refactored to \lstinline|const int BZ_RUNB=1;| in C, and later translated to \lstinline{const BZ_RUNB:i32=1;} in Rust. The refactor C code and the translated Rust code would both keep using the variable \lstinline{BZ_RUNB} instead of the value \lstinline{1}, while C2Rust directly puts the value into the translated code, lowing code readability.
Figure \ref{lst:C_const} shows these macros rewritten as constant global variables in C, and Figure \ref{lst:Rust_const} shows them rewritten as constant global variables in Rust.

\begin{figure}[!ht]
    \centering
    \begin{subfigure}[b]{.3\columnwidth}
        \begin{lstlisting}[language=C]
#define BZ_RUNB 1
#define BZ_M_IDLE 1
#define BZ_S_OUTPUT 1
        \end{lstlisting}
        \caption{Sample Numerical Macros in C}
        \label{lst:C_num_macros}
    \end{subfigure}%
    \hfill
    \begin{subfigure}[b]{.33\columnwidth}
        \begin{lstlisting}[language=C]
const int BZ_RUNB=1;
const int BZ_M_IDLE=1;
const int BZ_S_OUTPUT=1;
        \end{lstlisting}
        \caption{Rewritten as const global variables in C}
        \label{lst:C_const}
    \end{subfigure}
\hfill
\begin{subfigure}[b]{.33\columnwidth}
    \begin{lstlisting}[language=C]
const BZ_RUNB: i32=1;
const BZ_M_IDLE:i32=1;
const BZ_S_OUTPUT:i32=1;
        \end{lstlisting}
        \caption{Rewritten as global variables in Rust}
        \label{lst:Rust_const}
    \end{subfigure}

    \caption{Sample Translations of Numerical Macros}
    \label{fig:numerical_macros_definition}
\end{figure}

\paragraph{Replacing complex macros with functions.}
%

There can be complex macros, such as ``\lstinline|#define GET_MTF_VAL(label1,label2,lval) ...|'',  that accept arguments and implement complex logic and are reused many times in C code.
For better code readability and alignment with the original C code again, we rewrite the complex macros as the functions that have the same names as the macros by using LLM before applying our translation to the code using such macros.
Note that the arguments in complex macros are not typed and will be merely determined by the values passed to the macro when used.
Therefore,
when we prompt\footnote{\url{https://github.com/Cxm211/RustMap/blob/main/prompt-templates/macro-translation.txt}}
ChatGPT to rewrite the macro as a function, we also provide the code of the functions that use the macro, and instruct ChatGPT to replace the macro use with a function call if possible.
Taking the functions that use the macro as the context, ChatGPT is able to infer the data types of the macro arguments. Finally, the complex macros are replaced with the new functions, and the code using the macro is updated with the function call. The corresponding section\footref{example} in our repo provides a detailed example of complex macro handling.
Figure \ref{fig:complex_macro} shows an example, where one complex marco \lstinline{fswap(zz1, zz2)} is converted to a function \lstinline{fswap(UInt32* zz1, UInt32* zz2)}.

\begin{figure}[!ht]
    \centering

    \begin{subfigure}[b]{.45\columnwidth}
        \begin{lstlisting}[language=C, basicstyle=\tiny\ttfamily]
#define fswap(zz1, zz2) \
{ Int32 zztmp = zz1; zz1 = zz2; zz2 = zztmp; }




static
void fallbackQSort3 ( UInt32* fmap, 
                      UInt32* eclass,
                      Int32   loSt, 
                      Int32   hiSt )
{
  ...
  if (n == 0) { 
    fswap(fmap[unLo], fmap[ltLo]); 
    ltLo++; unLo++; 
    continue; 
  };
  ...
}
        \end{lstlisting}
        \caption{One Complex Macro in C}
        \label{lst:C_fun_macros}
    \end{subfigure}%
    \hfill
    \begin{subfigure}[b]{.45\columnwidth}
        \begin{lstlisting}[language=C, basicstyle=\tiny\ttfamily]
static void fswap(UInt32* zz1, UInt32* zz2) {
    UInt32 zztmp = *zz1;
    *zz1 = *zz2;
    *zz2 = zztmp;
}

static
void fallbackQSort3 ( UInt32* fmap, 
                      UInt32* eclass,
                      Int32   loSt, 
                      Int32   hiSt )
{
  ...
  if (n == 0) {
    fswap(&fmap[unLo], &fmap[ltLo]);
    ltLo++; unLo++;
    continue;
  }
  ...
}
        \end{lstlisting}
        \caption{Rewritten as a function in C}

        \label{lst:C_func}
    \end{subfigure}
    \caption{Preprocess of Complex Macros}
    \label{fig:complex_macro}
\end{figure}

Note that there are also macros that would correspond to syntactically invalid C code by itself, which are allowed by C but not allowed by Rust, and cannot be easily rewritten as simple variables or functions. It would be challenging to still retain such macros during translation, and we opted to apply the usual C preprocessor to expand such macros like other unhandled directives below.





\subsubsection{Unhandled conditional compilations and other directives}
\label{sec:unhandleddirectives}

Although many preprocessor directives can be translated by LLM, there are still many other directives (e.g., \lstinline|#if|, \lstinline|#pragma|, \lstinline|#line|, etc.) that may depend on system environment settings and require specialized handling and more advanced Rust macro features if to be retained and translated to Rust.
Moreover, many macro definitions inside \lstinline|#if| depend on system environment variables, which can only be determined how these macros are preprocessed at compile time.
One example of unhandled directives is provided in the examples section\footref{example} in our repo.
For example, \lstinline|#ifdef _WIN32| is often used to check whether the code is being compiled on a Windows system, allowing conditional compilation of platform-specific implementations, and the value of \lstinline|_WIN32| would be automatically defined by the compiler when targeting a Windows platform.

To ensure the generation of compilable and executable Rust programs when LLM fails to translate such diverse directives,
we use the same approach of other C-to-Rust translation tools, by preprocessing the directives using the GCC preprocessor to remove them or expand them into usual C code so that they can be translated with other code.
As a result, for each C file (\lstinline|.c| or \lstinline|.h|), we will have a preprocessed C file  (\lstinline|.i|) that contains only usual C code and the directives that can be translated.
Such \lstinline|.i| files 
from the given C project can still be compiled, executed with test cases, and 
will be analyzed by \ourapp further to identify dependencies and translation units.


\subsection{Function Dependencies and Translation Order}
\label{sec:dependency-order}


Project-scale C programs can contain numerous functions to be translated, necessitating a strategic prioritization of their translations.
Intuitively, the functions that are called by other functions should be translated before their callers;
and data types used by a function should be translated before the function.
If there are circular dependencies (e.g., recursive functions and data types), they can be put into one translation unit to be translated together, as long as their combined size is within the length limit of LLM.
In addition, to help check functionality based on a given test case, functions that are executed should be prioritized before others that are not.

To achieve this goal, we utilize both dynamic and static call graphs of the C program preprocessed (see Section \ref{sec:preprocessing}) by our approach to determine the translation order of functions and their dependencies.
A dynamic call graph represents the function cal relations in a program, by recording the actual calling sequence when running the program in debug mode with one or more test cases.
A static call graph represents conservatively all possible call relations.\footnote{We used gprof \cite{gprof} to get dynamic call graphs, and cflow \cite{cflowgnu} to get function call graphs and data type dependencies.}
While dynamic call graphs are more precise than static ones so that we can prioritize the translation of the executed functions with respect to given test cases as the Rust program can still compile and run even if other non-executed code is removed, static ones capture more call relations so that our translation for the program can be more complete in case more test cases will be used.\footnote{Due to this translation strategy, \ourapp cannot ensure the functional correctness of non-executed functions yet.}

We also note that there can be functions that call each other directly or indirectly, similarly for recursive data type definitions. For such cases, translating either function alone without referring to others may lead to translation errors
because LLM would lack the knowledge about the dependent code.
This motivates us to put inter-dependent functions and/or data types together for translation.
To do so, we use the Tarjan algorithm \cite{tarjan1972depth} to identify Strongly Connected Components (SCCs) in a call graph and their dependencies, and treat all the code belonging to one SCC as a translation unit.\footnote{If the number of lines of code of an SCC is beyond the length limit of LLM, we still break down it into smaller units for translation.}
%

\begin{figure}[h]
    \centering
    \includegraphics[width=0.5\columnwidth]{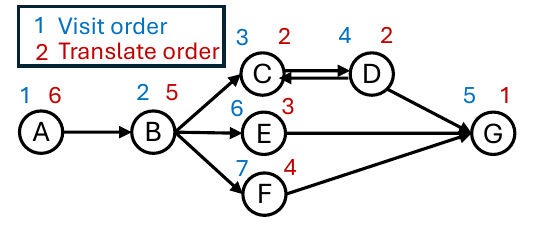}
    \caption{Illustration of translation order using a call graph. Each node represents a function. The blue numbers indicate the visit order, and the red numbers indicate the order of translation.}
    \label{fig:cg}
\end{figure}

Thus, to determine the translation units and their translation order, we start from the \lstinline{main} entry of a call graph of the project, we perform a depth-first search, recursively visiting all nodes and translating all the children of a node before translating the node itself, ensuring that each function is translated after all its callees and dependencies have been translated. 
Figure \ref{fig:cg} with colored numbers illustrates the visit order and the resulting translation units and order. E.g., Nodes C and D have the same colored ``Translate order'' $2$, which indicates they are in the 2nd unit to be translated together.

During this process, we have observed and utilized an important capability and flexibility of LLM in dealing with unknown function or data type: we can instruct LLM to ignore an unknown dependency when the C code for translation contains references to functions or data types \emph{not} defined in the code itself, or to replace the unknown with a generic no-op placeholder, or to ask human users to provide additional definitions of those dependencies. This is also a reason why we can just translate the functions that appear in dynamic call graphs to produce compilable and executable Rust program without the need to translate all functions in static call graphs.\footnote{This capability and flexibility of LLM is also applicable to any reasonable code chunk, not limited to a complete function. Thus, when the number of lines of code of just one function is beyond the length limit of LLM, we can break down the function into shorter code chunks at various syntactically valid places, e.g., the call sites to other functions, the boundaries of a \lstinline|while| statement, a \lstinline|case| in a \lstinline|switch| statement, etc., and treat each smaller code chunk as a translation unit and merge them back to a function after they are translated into Rust.}

\section{Translating C functions to Rust using LLM}
\label{sec:translation}

\begin{figure}[h]
    \centering
    \includegraphics[width=0.99\linewidth]{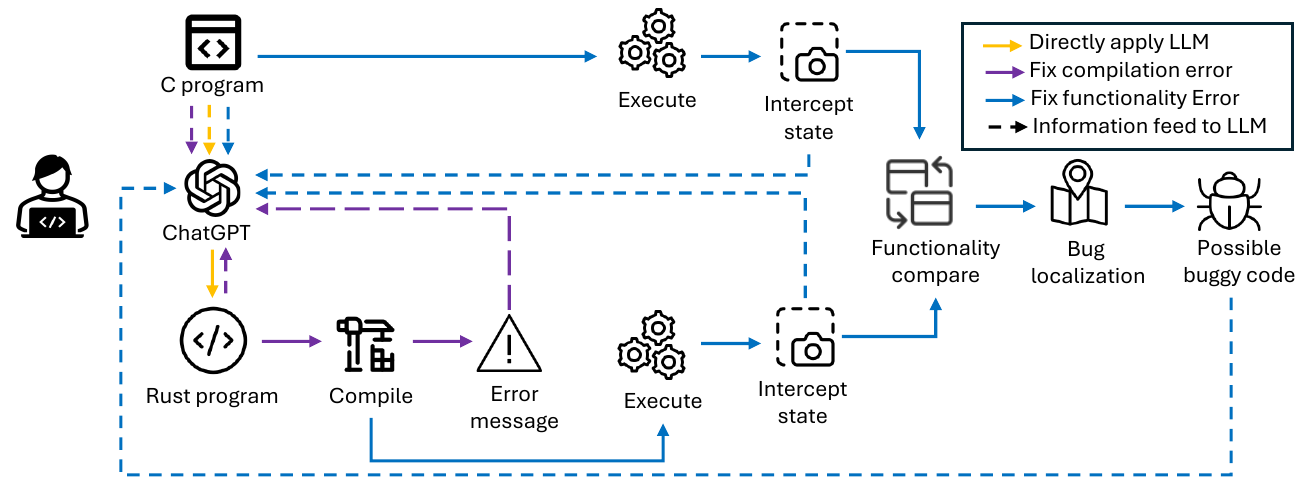}
    \caption{Workflow of translating a C program via LLM}
    \label{fig:synthesis-fn-level}
\end{figure}

Translating C programs to Rust programs needs to ensure that the Rust programs (1) can be compiled successfully, and (2) should behave consistently with the C programs.
We leverage ChatGPT 
4o \cite{gpt4o,openai2024gpt4}
with human prompts and test cases for these purposes.
Each program is broken down into translation units as described in the previous section; each unit is fed to LLM for translation, and their translation results are merged together to form the Rust program.

Figure~\ref{fig:synthesis-fn-level} shows an overview of our uses of LLM.
Specifically, given a C program and its translation units, we first prompt ChatGPT to translate the program (if the whole program can fit within the LLM input length limit) or one of its translation units (Section \ref{sec:chatgpt-translation}).
If the Rust compiler fails to compile the Rust code translated by ChatGPT, we will include the compiler error messages in the prompt to ChatGPT for the next round of trial (Section \ref{sec:compilation-errors}).
If the program and all its translation units can be translated and compiled, we continue to execute the translated program in the debug mode and check their execution states.
If there is a functional inconsistency between the Rust program and the C program when they execute, we 
perform bug localization to locate possible buggy code in the translated Rust code and its corresponding C code, and then prompt
ChatGPT to re-translate the code to fix the bug 
(Section \ref{sec:error-resolution}). The prompt templates we use are provided in our repo.\footnote{\label{prompttemplates} \url{https://github.com/Cxm211/RustMap/tree/main/prompt-templates}}

\subsection{Leveraging ChatGPT for Translation}
\label{sec:chatgpt-translation}

Rust and C are two distinct languages with different syntax and semantics.
This requires extensive pre-existing knowledge of both Rust and C, which is demanding and the main reason for translation errors by other tools. 
To address such issues, we utilize ChatGPT for translation, 
as ChatGPT is pre-trained with a large amount of programming-related knowledge, including C and Rust programs, which equips it with a strong ability to understand both languages.

Figure \ref{lst:prompt1} shows a prompt template\footref{prompttemplates} we used to translate a piece of C code to Rust.
The prompt considers various Rust programming conventions useful for idiomatic Rust code, e.g., distinguishing between mutable and immutable references, adding lifetime annotations, and using the Rust standard library as much as possible.
By following these conventions, the generated Rust code is more likely to be idiomatic and readable, and adhere to Rust's safety guarantees.

\begin{figure}[t]
    \begin{mdframed}
        \small
        Please take on the role of an expert developer familiar with the Rust and C and C++ programming languages.
        
        \smallskip
        Convert the given code to idiomatic Rust, keeping its functionality. Use minimal unsafe traits and follow the requirements below. Don't translate unknown variables or functions, and avoid assumptions. 
        If the C code references other vital functions or structures, ask me first and wait for my provided input. (ASK ME first)

        \begin{enumerate}[leftmargin=1.5em]
            \item If a variable inside the function is modified, add the \texttt{mut} specifier.
            \item Distinguish between mutable and immutable references by storing intermediate values.
            \item If necessary, add lifetime annotations.
            \item Add clear comments for all numeric types and pay attention to type conversions, especially between \texttt{usize} and others.
            \item Be cautious of potential out-of-bound errors in the C code.
            \item Use the Rust standard library as much as possible.
            \item When performing arithmetic operations, be mindful of potential overflow or underflow.
        \end{enumerate}
        \textbf{(...Here is C Code to be translated...)}
%
    \end{mdframed}
    \caption{Prompt for applying ChatGPT to translate C code to Rust}
    \label{lst:prompt1}
\end{figure}

\subsection{Adding Dependencies into Translated Files}
\label{sec:add-dependencies}

As our approach can divide a C file into multiple Rust files (cf.\ Section \ref{sec:structuring}) during translation,
we need to spell out the dependencies between the files so that the entire translated Rust project can be compiled successfully.
We identify the dependencies based on the original C code and add reference relations into the files.
Specifically, given a Rust file containing a translated function, we first identify all its callee functions using call graphs.
Then, we locate the corresponding Rust files for the callees and identify their module name.
Finally, we add Rust \lstinline{use} statements to import those modules at the beginning of the initial Rust file.
For example, \lstinline{BZ2_decompress} function calls \lstinline{BZ2_hbCreateDecodeTables} function, which is defined in \lstinline{huffman.c}.
We use the file path from the scaffolding step
as the module name (i.e., \lstinline{huffman::BZ2_hbCreateDecodeTables}), and import the module using \lstinline{use} statement at the beginning of the file containing the \lstinline{BZ2_decompress} function.
When some dependent functions are put into the same translation unit and their translation results are merged into one Rust file, the relevant \lstinline|use| statements would be removed.
This restructuring maintains the original dependencies between functions, enhancing the maintainability and readability of the translated Rust codebase.

\subsection{Fixing compilation errors}
\label{sec:compilation-errors}
We adapt a prior study \cite{rustassistance} that uses LLMs to produce fixes for compilation errors in Rust programs,
which achieves approximately 74\% peak accuracy by combining prompting techniques and iterative feedbacks.
When a function and its necessary dependencies are translated,
we use the \lstinline|rustc| compiler to check whether they can be compiled successfully.
If compilation fails, we will locate the Rust function or code construct that causes the failure, and take the compilation error messages, along with the buggy Rust code and its corresponding original C code, as the input prompt for ChatGPT to fix the errors in the Rust code.
We will conduct this process \edits{three times} until the generated Rust code can be compiled without an error.
\subsection{Locating and Fixing functionality errors}
\label{sec:error-resolution}
When the \lstinline|main| entry function and its necessary dependencies are translated and the whole Rust program is compiled,
we run the compiled Rust program with the test inputs available from the original C project.
To ensure they produce the same outputs for the same test inputs, we instrument the Rust program and intercept its runtime execution states, 
and then compare the states with those from the C program.\footnote{In our implementation, we inserted various \lstinline{print} statements at needed locations in both C and Rust programs to print out \emph{relevant inputs and outputs} and compared the printed values. Relevant inputs of a function are the function parameters; Relevant outputs of a function are the variables or expressions returned by the function. Relevant inputs of any code fragment are selected by LLM with a simple use-def static analysis of the data \emph{used} in the code fragment; Relevant outputs of any code fragment are selected by LLM with a simple use-def static analysis of the data that is \emph{modified} in the code fragment.}
If there is any inconsistency, we follow the steps below to {\bf locate} and {\bf fix} the possible faulty code fragments or functions.

To effectively {\bf locate} bug locations causing runtime state inconsistencies in programs, we adopt a binary-search-based method
on the C and Rust programs at the same time.
First, we consider all the statements in the buggy Rust program as bug candidate statements.
Then, a breakpoint in the Rust program is set around the halfway point of the sequential execution of all bug candidate statements; the breakpoint in the C program is set at the corresponding C statement semantically equivalent to the Rust breakpoint statement.
Then, we check the state of the Rust program and the state of the C program at the breakpoints.
If they are consistent, the bug candidate statements should be narrowed to its second half; otherwise, we narrow the bug candidate statements into its first half only (although the second half may still contain more bugs).
This process will be executed repeatedly until the number of bug candidate statements is small (fewer than 10 within a function) and the statements are relatively simple without complex control-flow changes (e.g., simple code blocks with less than 5 Cyclomatic Complexity \cite{cyclomatic_complexity,complexity_crate}).
As a result, we locate a code fragment in Rust with bug candidate statements (also called faulty code fragment), as well as a corresponding code fragment in the original C program that is supposed to be functionally equivalent to the Rust code fragment.

To {\bf fix} a possible faulty code fragment, we instantiate the prompt template shown in Figure \ref{lst:prompt3} \footnote{\url{https://github.com/Cxm211/RustMap/blob/main/prompt-templates/prompt_to_solve_inconsistency_error.txt}} and ask ChatGPT to fix.
We take (1) the corresponding code fragment from the C program with the states before and after executing the C code, and (2) the faulty Rust code fragment with the state before executing the faulty code, as the input prompt for ChatGPT to correct the Rust code so that it produces the same output state as the C code.
\begin{figure}[t]

    \begin{lstlisting}[frame=single, basicstyle=\tiny\ttfamily]
// C code fragment with its before- and after-states:
{
    (before-state of C)
    (...Here is the C code to be translated to Rust...)
    (after-state of C)
}

// Rust code generation:
{
    (before-state in Rust)
    (...Here is the possibily faulty Rust code...)
    /** please fix the Rust code fragment here to have
        consistent states as the C code above
      */
}
\end{lstlisting}
    \caption{Prompt template for ChatGPT to resolve functionality errors. The instantiated prompt will include the same requirements as shown in the prompt in Figure \ref{lst:prompt1}.}
    \label{lst:prompt3}
\end{figure}
This fix process for each located buggy code fragment will be iterated \edits{three} times. If ChatGPT fails to fix the Rust code to produce states consistent with the C code, 
we leave it as a translation failure case.
The whole locate-and-fix process will also be repeated many times \edits{until time out or the whole Rust program can run the given test cases to produce outputs consistent with the C program.}

\section{Empirical Evaluation}
\label{sec:evaluation}

To evaluate the effectiveness of \ourapp in translating C code to Rust, we perform studies on real-world C programs, including a set of RosettaCode programs 
and a file compression program {\tt bzip2}, ranging from tens of lines to tens of thousands of lines of code.
Section \ref{sec:trans-result-summary} shows the overall performance of \ourapp in various functionality, safety, readability metrics.
Section \ref{sec:advantages} provides translation examples (e.g., thread-safe global variables, 
interdependent data structures, pointer aliasing, control-flow \lstinline|switch-case|, \lstinline|goto|, etc.) that are challenging for other tools.
These examples are also provided in our repo\footnote{\url{https://github.com/Cxm211/RustMap/tree/main/case_study}} for potential reuses by future research.
Section \ref{sec:threats} discusses our limitations.

\subsection{Effectiveness}
\label{sec:trans-result-summary}

\subsubsection{Functionality Evaluation.}
\label{sec:functionality-evaluation}

After applying \ourapp, we successfully converted 125 RosettaCode programs and a file compression program {\tt bzip2} from C to Rust.
Unlike prior studies that only output Rust code snippets, we generated compilable Rust codebases that can execute and produce the same outputs as the original C programs for their given test cases.\todo{better to say what the test cases are / how the test cases were obtained, or provide the links to the test cases in the repo.}
For example, the converted {\tt bzip2} program can successfully compress and decompress given test files, while none of the existing tools can achieve this level of functionality.\footnote{The translated Rust code for bzip2 is at: \url{https://github.com/Cxm211/RustMap/tree/main/rust-code/bzip2_rs_gpt}.
\\
The translated Rust code for RosettaCode programs is at:
\url{https://github.com/Cxm211/RustMap/tree/main/rust-code/rosetta_code_gpt/125-rosetta-code-gpt}
}

On the other hand, there can be some C code that is not translated by our approach, as the C code can contain some functions, macros, or branches that are not executed by the given test cases.\todo{can be more convincing to say more about the test cases / how the test cases were obtained, and the link to some sample test cases that failed.}
To gauge the likely deficit, Table \ref{tab:merged_coverage_comparison} shows the proportion of C functions and lines that are translated into Rust {\it with respect to} the C functions and lines that are executed/covered by the test cases.

\begin{table}[ht]
    \centering
    \small
    \begin{tabular}{@{}lcc@{}}
        \toprule
        \textbf{C File} & \textbf{Function Coverage} & \textbf{Line Coverage} \\
        \toprule
        \multicolumn{3}{c}{{\tt bzip2}}                                              \\
        \midrule
        blocksort.c   & 66.7\%  & 71.7\%  \\
        bzip2.c       & 100.0\% & 100.0\% \\
        bzlib.c       & 100.0\% & 93.3\%  \\
        compress.c    & 100.0\% & 100.0\% \\
        decompress.c  & 100.0\% & 93.2\%  \\
        huffman.c     & 100.0\% & 98.8\%  \\
        \textbf{Average} & \textbf{94.4\%} & \textbf{92.8\%} \\
        \toprule
        \multicolumn{3}{c}{Rosetta Code} \\
        \midrule
        \textbf{125 .c files}   & \textbf{95.3\% }  & \textbf{93.2\% } \\
        \bottomrule
    \end{tabular}
    \caption{Proportion of C functions and lines that are translated into Rust {\it with respect to} the C functions and lines that are executed in the given test cases.}%
    \label{tab:merged_coverage_comparison}
\end{table}

We observe that the differences between the amount of C code that is translated and the amount of C code that is executed are small, indicating that the translated Rust projects have 
\edits{close}
functionality as the original C projects.

\rqbox{
\ourapp can produce runnable Rust projects that
pass the same test cases as the original C projects.
}

\subsubsection{Unsafety Evaluation.}
\label{sec:unsafety-evaluation}




We compare the number of unsafe code blocks in the Rust projects translated by \ourapp with those translated by other tools.
Laertes \cite{laertes21} implemented a tool to categorize unsafe code, and we used it for measuring unsafe code in Rust projects.
Table \ref{fig:Rosetta_Code} shows the numbers of unsafe code blocks in the Rust projects translated by \ourapp, C2Rust, CRustS, and Laertes.
We measure the unsafe code declared in Rust's official document \cite{rust_lang_unsafe_rust}: Read From Union (reading from a field of a C-style untagged union), MutGlobal Access (reading, writing, or referencing to mutable global(static)/external vars), RawPtr Deref (dereferencing a raw pointer), Alloc (direct external function calls to malloc and free), Extern Call (calling an external function other than `Alloc').

\rqbox{
\ourapp is much more effective in reducing unsafe code when translating C to functionally equivalent Rust.
}

\begin{table}[h]
    \centering
    \footnotesize
    {
    \begin{tabular}{rm{1cm}<{\centering}m{1cm}<{\centering}m{1cm}<{\centering}m{1cm}<{\centering}m{1cm}<{\centering}m{1cm}<{\centering}}
        \toprule
        \textbf{Tool} & \textbf{Read From Union} & \textbf{Mut Global Access} & \textbf{Raw Ptr Deref} & \textbf{Alloc} & \textbf{Extern Call} & \textbf{Unsafe Cast}   \\
        \toprule
        \multicolumn{7}{c}{{\tt bzip2}}                                              \\
        \midrule
        C2Rust             & 0                  & 386                & 3424            & 14             & 1               & 0               \\
        
        CRustS             & 0                  & 386                & 3424            & 14             & 1               & 0               \\
        
        Laertes            & 0                  & 386                & 3359            & 14             & 1               & 0               \\
        
        RustMap*           & 0                  & 126                & 122             & 5              & 0               & 0               \\
        \toprule
        \multicolumn{7}{c}{Rosetta Code}                                              \\
        \midrule
        C2Rust                & 0                  & 339                & 572             & 68             & 593             & 19              \\
        
        CRustS                & 0                  & 339                & 572             & 68             & 593             & 19              \\
        
        Laertes               & 0                  & 339                & 572             & 68             & 593             & 19              \\
        
        RustMap*              & 0                  & 3                  & 0               & 0              & 0               & 0               \\
        \bottomrule
    \end{tabular}
    }
    \caption{Numbers of unsafe code blocks}
    \label{fig:Rosetta_Code}
\end{table}

\subsubsection{Readability Assessment.}

\begin{figure}[h]
    \centering
    \begin{subfigure}[b]{0.4\linewidth}
    \includegraphics[width=0.7\textwidth]{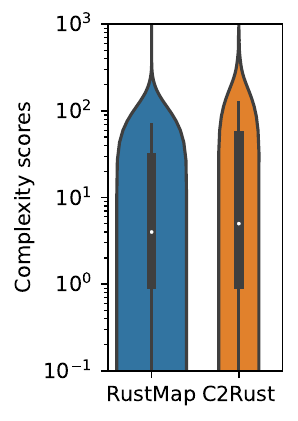}
    \caption{BZip2}\label{fig_f_bzip2_read}
  \end{subfigure}%
  \hspace{0.1\linewidth}%
  \begin{subfigure}[b]{0.4\linewidth}
    \includegraphics[width=0.7\textwidth]{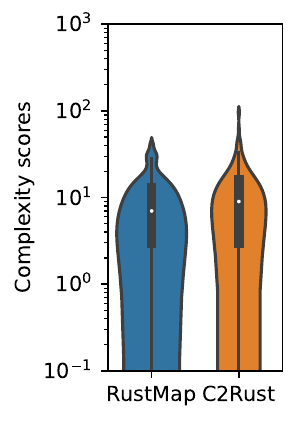}
    \caption{RosettaCode}\label{fig_f_rosetta_read}
  \end{subfigure}
    \caption{Cognitive Complexity of Rust code translated by \ourapp and C2Rust. Note that the y-axes are in {\bf log scale}.}
    \label{fig:readability}
\end{figure}

To measure \ourapp's capability in terms of code readability, we compare the \emph{Cognitive Complexity} (CC) scores of the Rust projects generated by different tools.
Cognitive Complexity quantifies code readability and maintainability, assessing the cognitive load on developers.
To do so, we apply the complexity crate \cite{complexity_crate} to the functions in the generated Rust projects.
Figure \ref{fig:readability} shows the comparison between \ourapp and C2Rust in terms of CC scores.
We observe that the complexity of the Rust projects translated by \ourapp is significant lower than the ones generated by C2Rust (p-value $<$ 0.05).

\rqbox{
Rust code translated by \ourapp is easier to read than the code translated by C2Rust.
}

\subsection{Advantages Against Other Tools: Case Studies}
\label{sec:advantages}



This subsection showcases various challenging examples during our translation of C programs to Rust. These sample cases demonstrate how \ourapp works better than other state-of-the-art C-to-Rust translation tools, including C2Rust \cite{c2rust_challenges}, CRustS \cite{crusts} and Laertes \cite{laertes21, laertes23}, and the translation patterns identified during the process can be generally applicable to all future C-to-Rust translation studies aiming for producing more readable and safer code.

\subsubsection{\textbf{Rewriting Global Variables}.}
\label{sec:lazy-static}
Global variables are widely used in C programs,
but if directly translated into Rust, they are not thread-safe and can cause static initialization order fiasco \cite{RustBookUnsafe,cppinitfiasco} if imported multiple times into different files. Thus, simple syntax-rule-based translation done by other tools are insufficient in ensuring the safety of the translated code.

For global variables that are of primitive types in C, 
we observe that we can still use simple rule-based automation translation like other tools without the cost of invoking LLMs and associated uncertainty and error fixing, but need to enforce the global variables are initialized only once in a thread-safe manner when they are first accessed by using the \lstinline{lazy_static} macro and \lstinline|Mutex| in Rust.
%
Specifically, we rewrite C primitive-type global variables using \lstinline{lazy_static} and providing getter and setter functions to replace direct accesses to them, as recommended for Rust programmers \cite{rust_forum_lazy_static}.
Figure \ref{fig:globalvar} illustrates sample global variables declared in C (Figure \ref{lst:declaration_global_C}) and their translations by C2Rust (Figure \ref{lst:C2Rust_global_declaration}) and by \ourapp (Figure \ref{fig:Rustmap_global_declaration}).
For example, we rewrite \lstinline{int numFileNames} via a \lstinline{Mutex} inside the body of the \lstinline{lazy_static} macro. In the Rust code in Figure \ref{fig:Rustmap_global_declaration}, \lstinline{lazy_static} is used to create lazily instantiated global variables.
This is necessary because Rust does not support mutable global variables directly. By using \lstinline{lazy_static} and \lstinline{Mutex}, we ensure that these global variables are safely initialized only once, and access to them is thread-safe, preventing data races.
This approach replaces the traditional global variables in C, offering better safety and concurrency guarantees. and define the \lstinline{get_numFileNames} and \lstinline{set_numFileNames} functions.
%
As a comparison, the code translated by C2Rust would lead to reading or modifying a mutable static variable, that is an unsafe practice in Rust \cite{RustBookUnsafe}.

\begin{figure}[!thb]
\begin{subfigure}{\columnwidth}
\begin{lstlisting}[frame=single, basicstyle=\scriptsize\ttfamily,language = C]
// bzip2.i
int numFileNames, numFileProcessed, blockSized100k;
\end{lstlisting}
\caption{Sample Global Variable Definitions in C}
\label{lst:declaration_global_C}
\end{subfigure}

\smallskip

\begin{subfigure}{\columnwidth}
\begin{lstlisting}[frame=single, basicstyle=\scriptsize\ttfamily,language = C]
// bzip2.rs from C2Rust
#[no_mangle]
pub static mut numFileNames: Int32 = 0;
#[no_mangle]
pub static mut numFilesProcessed: Int32 = 0;
#[no_mangle]
pub static mut blockSize100k: Int32 = 0;
\end{lstlisting}
\caption{Sample Global Variable Definitions translated by C2Rust}
\label{lst:C2Rust_global_declaration}
\end{subfigure}

\smallskip

\begin{subfigure}{\columnwidth}
\begin{lstlisting}[frame=single, basicstyle=\scriptsize\ttfamily,language = C]
// idiomatic lazily instantiated global variable
// bzip2/global_bzip2.rs
lazy_static! { 
    pub static ref numFileNames: Mutex<i32> = Mutex::new( 0 );  
    pub static ref numFilesProcessed: Mutex<i32> = Mutex::new( 0 );   
    pub static ref blockSize100k: Mutex<i32> = Mutex::new( 0 );
}

pub fn get_numFileNames() -> i32 { /.../  } 
pub fn set_numFileNames(new_value: i32) {  /.../ }

pub fn get_numFilesProcessed() -> i32 { /.../  } 
pub fn set_numFilesProcessed(new_value: i32) {  /.../ }

pub fn get_blockSize100k() -> i32 { /.../  } 
pub fn set_blockSize100k(new_value: i32) {  /.../ }
\end{lstlisting}
\caption{Sample Global Variables translated via \lstinline{lazy_static!} by \ourapp}
\label{fig:Rustmap_global_declaration}
\end{subfigure}

\caption{Sample Translations of Global Variables}
\label{fig:globalvar}
\end{figure}

For non-primitive global variables, because it is more challenging to use rule-based automation to rewrite accesses into various data fields of non-primitive global variables, we leave it to LLMs to translate them following the same translation process for other code constructs (see Section \ref{sec:chatgpt-translation}) and rely on the same error checking and fixing processes (see Sections \ref{sec:compilation-errors} and \ref{sec:error-resolution}) to ensure their functionality.\footnote{The functionality is thus only ensured with respect to the given test runs.}

\subsubsection{\textbf{Pointer Aliasing}.}
\label{sec:pointer-aliasing-example}
In C, pointer aliasing \cite{amd_pointer_aliasing}, especially involving arrays, is often used to optimize memory usage and access patterns. The flexibility to cast pointers from one type to another enables flexible manipulation of data structures with varied granularity. However, it may lead to potential issues with type safety and endianness sensitivity.

For example, in Figure \ref{fig:Pointer-Aliasing-in-C} excerpted from bzip2, the \texttt{EState} structure contains three array pointers of different data types, and a 32-bit signed integer that indicates the number of blocks.
The line 10 \lstinline{s->block = (UChar*)s->arr2;} casts \lstinline{UInt32*} to \lstinline{UChar*}; the resulting pointer \lstinline{block} is thus an alias of \lstinline{arr2} of a different type and can access individual bytes of the integers pointed to by \lstinline{arr2}.
Whether the memory operations done through \lstinline{block} is correct is dependent on the system's byte order/endianness. 

\textbf{Rewrite Pointer Aliasing in C2Rust.}
Figure \ref{fig:c2rust-duplicate-definition} shows the Rust code translated by C2Rust:
C2Rust keeps \lstinline{arr2}, \lstinline{block}, and \lstinline{zbits} as raw pointers, and uses \lstinline{unsafe} code blocks, that can increase the risk of unsafe memory operations and errors.

Another issue with the C2Rust translation is that it produces multiple definitions of \lstinline{struct EState} across different translated files \lstinline{compress.rs}, \lstinline{blocksort.rs}, and \lstinline{bzlib.rs}, even though the struct definition only appears once in one header file \lstinline{bzlib_private.h} in the original C program.
C2Rust and all C2Rust-based tools \cite{laertes21,laertes23,crown,crusts} have this issue because it preprocesses each \lstinline{.c} file into \lstinline{.i} first before translation and if multiple \lstinline{.c} files contain the same \lstinline{.h} file, the header contents would appear multiple times across the files and C2Rust would produce erroneously duplicate code.

\begin{figure}[ht]
\begin{subfigure}{\columnwidth}
\begin{lstlisting}[frame=single, basicstyle=\scriptsize\ttfamily,language = C]
// bzlib_private.h
typedef struct {
    UChar* zbits;
    UInt32* arr2;
    UChar* block;
    Int32  nblock;
} EState;
// pointer aliasing without endianness concern (in bzlib.c)
// initialize EState *s
s->block = (UChar*)s->arr2;
// pointer aliasing with endianness concern (in compress.c)
s->zbits = (UChar*) (&((UChar*)s->arr2)[s->nblock]);

// passing one of both fields into calling function in fallbackSort
fallbackSort (/* ... */ s->arr2, /* ... */);
otherFunction (s->zbits);
\end{lstlisting}
\caption{Sample \texttt{struct} with Pointer Aliasing in C.}
\label{fig:Pointer-Aliasing-in-C}
\end{subfigure}

\smallskip

\begin{subfigure}{\columnwidth}
\begin{lstlisting}[frame=single, basicstyle=\scriptsize\ttfamily,language = C]
// Duplicated Defined in compress.rs, blocksort.rs, bzlib.rs
pub struct EState {
    pub arr2: *mut UInt32,
    pub block: *mut UChar,
    pub zbits: *mut UChar,
    pub nblock: Int32
    /* ... */
}
// bzlib.rs (without endianness concern)
unsafe{ (*s).block = (*s).arr2 as *mut UChar; }
// compress.rs (with endianness concern)
unsafe {
    (*s).zbits =
        &mut *((*s).arr2 as *mut UChar).offset((*s).nblock as isize) as
            *mut UChar;
}

// blocksort.rs
fallbackSort(/* ... */, (*s).arr2, /* ... */);
\end{lstlisting}
\caption{Pointer Aliasing translated by C2Rust.}
\label{fig:c2rust-duplicate-definition}
\end{subfigure}

\caption{Sample Translations of Aliased Pointers for C and C2Rust}
\label{fig:aliased-pointer-C-and-C2Rust}
\end{figure}

\begin{figure}[!th]
\begin{subfigure}{\columnwidth}
\begin{lstlisting}[frame=single, basicstyle=\scriptsize\ttfamily,language = C]
// Define only once in headers/bzlib_private.rs 
// use bzlib/globals_bzlib.rs 
pub struct EState {
    pub block: Vec<u8>, // use block to represent zbits
    // pub arr2: Vec<u32> 
    // pub zbits: Vec<u8>,
    pub nblock: i32,
}
\end{lstlisting}
\caption{Rewrite \lstinline{struct} EState by \ourapp. The \lstinline{struct} translated by \ourapp is stored in a header file generated by the project scaffolding step (Section \ref{sec:scaffolding}) and shared with all other files needing this struct.}
\label{fig:rustmap-struct}
\end{subfigure}
\smallskip
\begin{subfigure}{\columnwidth}
\begin{lstlisting}[frame=single, basicstyle=\scriptsize\ttfamily,language = C]
pub fn rebuild_arr2_from_block(block: &[u8]) -> Vec<u32> {
    /* reconstructs a u32 array from a block of bytes */
}

pub fn build_block_from_arr2(arr2: &[u32]) -> Vec<u8> {
    /* converts a u32 array into a block of bytes */
}
// blocksort/scc_71_fallbacksort.rs
let mut arr2 = build_block_from_arr2(s.block);
fallback_sort(/*...*/, &mut arr2, /*...*/);
// Subsequent assignment of modified temporary arr2 to s.block
s.block = rebuild_arr2_from_block(&arr2);
\end{lstlisting}
\caption{Aliased Pointer Operations WITHOUT Endianness Concern.}
\label{fig:rustmap-ptr-aliasing}
\end{subfigure}

\smallskip

\begin{subfigure}{\columnwidth}
\begin{lstlisting}[frame=single, basicstyle=\scriptsize\ttfamily,language = C]
pub fn get_zbits_from_arr2(estate: &mut EState) {
    /* construct a temporary arr2:Vec<u32> from estate.block:Vec<u8> */
    /* updates the zbits field based on the arr2, and
       Uses if-else checks based on system endianness. E.g., */
    let mut bytes = if cfg!(target_endian = "little") {
        arr2[offset].to_le_bytes() // Convert to little-endian byte array
    } else {
        arr2[offset].to_be_bytes() // Convert to big-endian byte array
    };
}

pub fn update_arr2_from_zbits(s: &mut EState, zbits: &[u8]){
    /* updates the arr2 field based on the zbits values.
       Uses if-else checks based on system endianness. E.g., */
    arr2[offset] = if cfg!(target_endian = "little") {
        u32::from_le_bytes(bytes) // Convert bytes back to little-endian u32
    } else {
        u32::from_be_bytes(bytes) // Convert bytes back to big-endian u32
    };
    /* converts arr2 array into s.block */
}

// In BZ2_compressBlock
let mut s = // initialize new EState
let mut zbits: Vec<u8> = get_zbits_from_arr2(&mut s); // obtain zbits
otherFunction(&zbits, ...);
update_arr2_from_zbits(&mut s, &zbits); // update block
\end{lstlisting}
\caption{Aliased Pointer Operations WITH Endianness Concern.}
\label{fig:rustmap-ptr-aliasing-endianness}
\end{subfigure}

\caption{Pointer Aliasing Rewrites by \ourapp}
\label{fig:pointer-writes-rustmap}
\end{figure}

\textbf{Rewrite Pointer Aliasing in \ourapp.}
\ourapp replaces C pointer aliasing in a safer and idiomatic way that uses vectors and conversion functions.
First, when prompting LLM with the struct definition in C, we also ask LLM to infer the suitable vector type for each pointer when converting it to Rust.
For example, Figure \ref{fig:rustmap-struct} shows that the pointers are converted to use safe collections like \lstinline{Vec<u8>} and \lstinline{Vec<u32>}.
However, note that both \lstinline{arr2} and \lstinline{zbits} are commented out, because they point to the same memory as \lstinline{block} in the original C, and we can use just one vector to represent all three.
We also used LLM to identify potential aliasing among all pointers in the given code during prompting. 
Then, to mimic the memory operations done via aliased pointers in the original C program, we also prompt LLM to construct specific functions for each aliasing pointer to read from and write to the aliased memory.
Figure \ref{fig:rustmap-ptr-aliasing} and Figure \ref{fig:rustmap-ptr-aliasing-endianness} show-case such functions for the \lstinline{arr2} and \lstinline{zbits} pointers, respectively.
The internal workings of the functions vary, depending on whether they are concerned of endianness.

\paragraph{Case 1: Pointer Aliasing without Endianness Concern.}
In Figure \ref{fig:rustmap-ptr-aliasing}, the two conversion functions
are constructed by ChatGPT with human feedback. 
They essentially make copies of the array referenced by \lstinline{block} before/after each read/write operation for safer access.
Lines 8--12 illustrates their usage for migrating C code involving operations on aliasing pointer: a temporary variable \lstinline{arr2} is used to hold a new copy of the original array referenced by \lstinline{s.block}; then all read/write operations via the original \lstinline{arr2} in C are done via the temporary variable; finally, a new copy of the array referenced by the temporary variable is passed back to the original reference \lstinline{s.block}.
Tihs way of translation produces safer access to shared arrays, while it may still not be safe for multi-thread programs and slow down the efficiency of the code. We leave its improvements for future work.

\paragraph{Case 2: Pointer Aliasing with Endianness Concerns.}
Endianness concerns arise in C due to direct memory manipulation that may be sensitive to byte order, as shown in Figure \ref{fig:Pointer-Aliasing-in-C} line 12.
In this line, \lstinline{arr2} is first cast to \lstinline{UChar*}. Then, the address of the \lstinline{nblock}-th element of this array is taken and cast to \lstinline{UChar*} and assigned to \lstinline{zbits}. These seemingly convoluted operations may indicate the code's intention to be endianness-aware, and thus the translated Rust code should also ensure correct handling of endianness.

Figure \ref{fig:rustmap-ptr-aliasing-endianness} illustrates the two conversion functions and their usages for such cases. The essential idea of the conversion functions is the same as the case without concerning endianness by using a temporary variable to hold copies of the array; the main differences are the if-else checks of system endianness to manage byte order correctly when making copies of the array.

Compared to the direct raw pointer operation translated by C2Rust (Figure \ref{fig:c2rust-duplicate-definition}), \ourapp handles pointer aliasing and endianness in a much safer way.
Whether to be concerned of endianness when translating operations on an aliased pointer, we also prompted ChatGPT to make the choice. 

\rqbox{
When using \ourapp to translate pointer aliasing operations, we found that ChatGPT helps to provide safe rewrites for either without or with concerns of endianness \cite{ibm_endianess} (illustrated in Figure \ref{fig:rustmap-ptr-aliasing} and Figure \ref{fig:rustmap-ptr-aliasing-endianness} respectively). 
}

\subsubsection{\textbf{Control-Flow Translation: \texttt{switch-cases}}.}
\label{sec:switch-cases}

In C, \lstinline{switch} statements can fall through \cite{learncpp_switch_fallthrough_scoping}, i.e., the execution of a \lstinline{case} that does not contain a \lstinline{break} statement will continue directly into the next \lstinline{case}.
In contrast, Rust's \lstinline|match| statement does not have fall-through; each arm must explicitly end with a value or an action, ensuring that only one arm is executed.
Figure \ref{fig:C-control-flow} shows the original C code with a switch-case structure.
Figure \ref{fig:C2Rust-control-flow} displays the Rust code rewritten by C2Rust.
%
In contrast, Figure \ref{fig:RustMap-control-flow} shows the Rust code translated by \ourapp, as suggested by ChatGPT.
By leveraging match statements in a loop that encapsulates state transitions explicitly,
it reduces unsafe code usage, and improves readability and maintainability.

\begin{figure*}[ht]
    \centering
    
    \begin{subfigure}[b]{.32\textwidth}
\begin{lstlisting}[ basicstyle=\tiny\ttfamily]
// decompress.i 
switch (s->state) {
    /* ... */ 
    case 11:
    /* code without break */
    case 12:
    /* code without break */
    /* ... */
}
        \end{lstlisting}
        \caption{Sample C \texttt{switch-cases} with fallthrough}
        \label{fig:C-control-flow}
    \end{subfigure}%
    \hfill
    \begin{subfigure}[b]{.32\textwidth}
\begin{lstlisting}[basicstyle=\tiny\ttfamily]
match (*s).state {
    /* .. ... */
    11 => current_block
          = 12259750428863723923,
    12 => current_block
          = 15146946972525368609
    /* ... */ 
}
match current_block {
    12259750428863723923 => {
         /* ... */ 
    }
    15146946972525368609 => {
         /* ... */  
    }
     /* ... */
}
        \end{lstlisting}
        \caption{Rewrite by C2Rust}\todo{something is missing for the match statement}
        \label{fig:C2Rust-control-flow}
    \end{subfigure}%
    \hfill
    \begin{subfigure}[b]{.32\textwidth}
\begin{lstlisting}[ basicstyle=\tiny\ttfamily]
// decompress/scc_59_BZ2_
decompress.rs
'state_loop: while s.state <= 50 {
    match s.state {
        /* ... */
        11 => {
            /* ... */
            s.state = 12
        },       
        12 => {
            /* ... */
            s.state = 12
        }
        /* ... */
    }
}
        \end{lstlisting}
        \caption{Rewrite by \ourapp.}
        \label{fig:RustMap-control-flow}
    \end{subfigure}
    \caption{Sample Translations of {\tt switch-cases} with fallthrough.}
    \label{fig:switch-cases}
\end{figure*}

\subsubsection{\textbf{Control-Flow Translation: \texttt{goto}}.}
\label{sec:goto}

\lstinline{goto} statements, commonly used in C for jumping to specific code locations, are not support in Rust.
Figure \ref{lst:goto-case-study-in-C} is an example of using \lstinline{goto} in C code.
The \lstinline{GET_BITS} macro calls another macro, \lstinline{RETURN}, that utilizes \lstinline{goto}. \lstinline{GET_BITS} itself is used within a \lstinline{switch-case} with fallthrough.
When prompted with all the code at once, ChatGPT could not generate correct translated Rust, although it provided valid suggestions to consider wrapping the code at the goto target into a function and splitting the code in the macro into two parts at the conditional statement checking for the \lstinline{goto}.
We then refined the prompts for ChatGPT according to its own suggestions, to (1) wrap the code at and after the \lstinline{save_state_and_return} target label as a function and replace the goto with a function call, and (2) split the code at the if statement and replace the use of the \lstinline{GET_BITS} macro with calls to the two split functions.
ChatGPT was then successful in performing the tasks, producing one function \lstinline{save_state_and_return} for the goto and two split functions \lstinline{GET_BITS_first_half} and \lstinline{GET_BITS_second_half}, based on its understanding of the functionality of the code.
\begin{itemize}[nosep,leftmargin=1em]
\item In the C code, the \lstinline{GET_BITS} macro intends to extract a specified number of bits (\lstinline{nnn}) from the input stream and stores them in the variable \lstinline{vvv}.
\item When there are not enough available bits in the input stream, the \lstinline{RETURN} macro is invoked to save the current state and use \lstinline{goto} to jump to the \lstinline{save_state_and_return} label. 
\end{itemize}
The first split function attempts to perform the same functionality of the code before the if statement and returns a \lstinline{Result} to indicate if it is successful (\lstinline|OK|) or not (\lstinline|Err|). 
The second split function wraps all the remaining code after the if statement, ensuring correct functionality.
Figure \ref{lst:goto-case-study-in-Rust} shows the Rust code by \ourapp.
This splitting and replacement worked well, although specific for this case; it makes the translated code more readable and maintainable through explicit error handling and state management.

In contrast, C2Rust uses a more generic approach to compile away \lstinline{goto} and \lstinline{switch-cases} by implementing the Relooper control-flow analysis and restructuring \cite{Ramsey2022relooper,peterson1973capabilities,Bahmann2015ddg}, 
but it does not 
consider project-specific code functionality, and its translated code grows much in size and becomes harder to read.
Future work may consider using Rust \lstinline|goto| macros \cite{crate-goto} with the prompts of \ourapp to make the translated code more readable while retaining code functionality.

\begin{figure}[th]
\begin{subfigure}{\columnwidth}
\begin{lstlisting}[frame=single, basicstyle=\scriptsize\ttfamily,language = C]
// definition of two macros
#define GET_BITS(lll,vvv,nnn)                     \
  ...                                             \
  if (s->strm->avail_in == 0) RETURN(BZ_OK);      \
  ...

#define RETURN(rrr)                               \
   { retVal = rrr; goto save_state_and_return; };

// usage of two macros
switch (s->state) {
  // case 13:
  GET_BITS(BZ_X_MAGIC_4, s->blockSize100k, 8)
\end{lstlisting}
\caption{Sample goto in C}
\label{lst:goto-case-study-in-C}
\end{subfigure}

\begin{subfigure}{\columnwidth}
\begin{lstlisting}[frame=single, basicstyle=\scriptsize\ttfamily,language = C]
fn save_state_and_return (s: &mut DState, /* ... */ ) {
    /* Code at and after goto target */
}
...
match s.state {
 13 => {
// #[cfg(debug_assertions)]
  s.state = 13;
  loop {
    let mut tmp_blockSize100k =  s.blockSize100k as u32;
    match GET_BITS_first_half(s, &mut tmp_blockSize100k, 8){     
        Ok(_) => {  s.blockSize100k = tmp_blockSize100k as i32; break;  }
        Err(_) => { s.blockSize100k = tmp_blockSize100k as i32; }, 
    } 
    if unsafe { (*s.strmD).avail_in } == 0 {  
        retVal = CONSTS.BZ_OK;   save_state_and_return(s, /* ... */);
        break; // break out the inner loop
    }
    GET_BITS_second_half(s);
...
\end{lstlisting}
\caption{Sample Rust without \lstinline{goto} translated by \ourapp.}
\label{lst:goto-case-study-in-Rust}
\end{subfigure}

\caption{Sample \lstinline{goto} inside macro definitions.}
\label{fig:goto+macro}
\end{figure}

\subsection{Discussion \& Threats to Validity}
\label{sec:threats}
There are still different kinds of code features in C imposing challenges when being translated to Rust.
For example, 
C projects often involve sharing arrays by pointers, while \emph{safe} Rust disallows two writable pointers referencing the same memory to avoid memory errors. \ourapp translates array pointers to Rust vectors (\texttt{Vec<>}) and enable vector sharing by making value copies of the vectors before and after each write operation, which may impact code efficiency, instead of using slices to improve efficiency.
Managing lifetimes of slices \cite{rust_lifetime_2023} in Rust can be error-prone and challenging, which might impact performance and correctness.
Pointer punning \cite{wikipedia_type_punning} and aliasing are very common in C projects
within the same struct, 
where two array pointers point to the same memory, while \emph{safe} Rust disallows two writable pointers referencing the same memory to avoid memory errors.
\ourapp translates these pointers into \texttt{Vec<>} and uses value copying to handle read/write operations of the data, which is much less efficient than original C.
Using slices within the same struct introduces lifetime issues \cite{rust_lifetime_2023}, making this approach worth finding solutions to resolve. This is because slices are faster than vectors in Rust. Slices are references to existing data and do not involve the overhead of memory allocation and deallocation, resulting in more efficient operations.
Future work could explore the deeper semantic of such pointer sharing operations (e.g., there may be at most one writer to the shared memory at any time) to enable more efficient translation to more idiomatic Rust code, 
although this may involve trade-offs in machine performance.

Concurrent programming in C also often involves shared memory, which has very different coding patterns and algorithms from Rust. Our study has not yet evaluated with C code having concurrency and threads, and future work is to leverage Rust's safe concurrency features during translation to ensure idiomatic coding and correct multi-threading capabilities of translated code.

C code can manipulate memory at byte level and becomes sensitive to the endianness/byte order in the execution systems.
\ourapp can infer the expected endianness of the code based on LLM prompts and add suitable checks (e.g., \lstinline{cfg!(target_endian = "little")}) in the translated Rust code as needed. 

A trivial but important aspect of C-to-Rust translation involves variable naming conventions. E.g., in C, camel case is commonly used, whereas Rust developers may prefer underscores or snake cases.
Thus, during the conversion, variable names should be transformed from camel case to snake case to adhere to Rust's naming conventions.
ChatGPT used in \ourapp has the capability to rename variable names as prompted.

\ourapp focuses on translating functions executed by given test cases. We only used a few test cases for each of the programs in our evaluation, and the code coverages of the original C programs have not reached 100\%. But we ensure that for the given test cases, the translated Rust code, can function correctly in correspondence to the test cases.

Future initiatives could apply \ourapp to more C projects, to explore the promise and peril of using LLMs, as there are so many interesting but different code features in C and Rust that may benefit from LLM-based translation.

Various steps in \ourapp involve customized prompts to deal with different code patterns and translation errors, which involved much manual effort in developing the prompt templates and deciding which prompts to use for different situations. Future efforts could automate and systematize the prompting process better involving various LLM-based agents, especially for large-scale projects.
This would involve developing frameworks and tools to facilitate the efficient and accurate translation of extensive C codebases. 
An integration and validation phase using static analysis and testing tools to evaluate C and Rust code complexity and functionality could also be beneficial.
For intricate code segments, symbolic execution might be employed for preemptive issue resolution. Additionally, enhancing ChatGPT's role in translating C to Rust and improving test coverage could streamline the process, ensuring the accuracy and functionality of the converted code.

Further efforts could also explore improving the speed of array pointer aliasing, vector copying, and restoration through multithreading, despite potential trade-offs in machine performance. This enhancement aims to optimize the efficiency of handling arrays and vectors in the translated code, mitigating some performance issues encountered in the current conversion process.

\section{Related work}
\label{sec:related}
\todo{add a few more recent related work: 
not published: 
"SmartC2Rust: Iterative, Feedback-Driven C-to-Rust Translation via Large Language Models for Safety and Equivalence"; 
LLMigrate: Transforming ``Lazy'' Large Language Models into Efficient Source Code Migrators;
}
Unsafe Rust code has been used extensively~\cite{howunsaferust20} by developers, as it is difficult to write safe code that implements complex and efficient operations under the restrictions of the Rust type system. 
Efforts to reduce the amount of unsafe code in Rust, in the context of Rust translated using C2Rust, have been diverse and significant.
CRustS~\cite{crusts} utilizes TXL source-to-source transformation rules to minimize the extent of unsafe expressions in the translated Rust code. Laertes~\cite{laertes21,laertes23} concentrates on diminishing unsafe code by addressing import and lifetime errors related to raw pointers, a common issue in translations from C. Meanwhile, Crown~\cite{crown} offers a suite of analyses aimed at inferring the ownership models of C pointers, facilitating their translation into safe Rust equivalents.
Hong et al. \cite{hong2024don} used a static analysis based on abstract interpretation, complemented by the concept of abstract read/write sets, to replace output parameters (i.e., pointer-type parameters for producing outputs) in C programs with `Option/Result` (the return type of a function that represents functions returning multiple values and may fail) in the translated Rust code.
Hong et al. \cite{hong2024tag} further developed static analysis to identify tag fields and values, using must-points-to analysis and heuristics, to replace unions with tagged unions during C-to-Rust translation.
Further, Hong et al. \cite{hong2025type} proposed tools to port entire C programs to Rust by translating each C function to a Rust function with a signature containing proper Rust types through type migration with the help of LLM, 
but they have not been able to resolve all type errors to produce executable Rust programs.

LLMs have also been applied in different ways to help translate programs to Rust. Zhang et al.\ \cite{zhang2024scalable} utilize type compatibility and feature mapping between source and target languages to facilitate translation and have achieved meaningful results when translating Go code to Rust.
Eniser et al.\ \cite{eniser2024towards} evaluate various LLM models on translating a few Go and C programs to Rust; they use differential fuzzing to check if a Rust translation is I/O equivalent to the original source program and use feedback mechanisms to repair faulty translation. They achieved a translation success rate of about 47\% and provided insights for improvements.
Although we share a similar high-level idea of using LLMs for translation, our approach uses different code break-down strategy and different mechanisms to check code correctness and provide feedback for LLMs. It would be valuable future work to explore and compare different strategies in using LLMs and agent workflows to improve the translation results further.

In the realm of Rust synthesis, there have been innovative attempts to enhance safety and correctness. Synthetic Ownership Logic (SOL)~\cite{lev23} aims to provide correct-by-construction programs in safe Rust. Furthermore, the program synthesis for Rust library API testing\cite{SyRust2021} explores the synthesis of safe Rust code while dealing with unsafe code.

Regarding the verification of Unsafe Rust, notable advancements have been made. RustBelt~\cite{rustbelt17} offers the first formal (and machine-checked) safety proof for Rust. Building on this, RustHornBelt~\cite{rusthornbelt22} extends these proofs to ensure the soundness of Rust APIs implemented with unsafe Rust code. Recently, \cite{commonbugFixRust} has studied on how to implement a novel embedding of Rust code into fixed-sized vectors to categorize and find Rust common fix patterns.

Recent research covers a range of topics: \cite{rust_uf} explores the impact of unstable features on package compilation, \cite{rust_rpg} presents a technique for automatic fuzz target synthesis in Rust libraries, \cite{rust_unsafe} studies safety requirements in unsafe Rust code, \cite{rust_oss} examines differences between paid and volunteer developers in the Rust community, \cite{rust_lancet} introduces a tool for fixing ownership-rule violations, \cite{rust_clippy} analyzes the effectiveness of Clippy lints, \cite{rust_security} investigates security vulnerabilities in Rust packages, and \cite{rust_klee} demonstrates the use of KLEE for detecting unrecoverable errors. These papers provide significant insights and tools that contribute to the security, reliability, and development practices of the Rust ecosystem, while targeting different problems from our C-to-Rust code translation problem.

\section{Conclusion \& Future Work}
\label{sec:conclusion}

Unlike previous translation tools such as C2Rust, which focused primarily on syntactic conversions without considering functional integrity, our approach, \ourapp, with large language models (LLM) and program analysis, 
shows much potential in performing more semantic-aware, project-scale code translation.
We have successfully translated 125 RosettaCode programs and the bzip2 file compression program into Rust, enhancing the functional usability of the translated Rust code with better code safety and readability. 

Looking ahead, we plan to expand our project scope to include more comprehensive translations from C to Rust, 
to further explore the capabilities of LLMs in aiding code conversion of diverse features and to conduct more performance evaluations. Our long-term aim is to ensure not only the functionality, readability, safety, but also performance of translated code and their compatibility with untranslated code,
fulfilling the promise of Rust to replace C as a better programming language. 



%
%
%
\bibliographystyle{splncs04}
\bibliography{main}

\begin{thebibliography}{10}
\providecommand{\url}[1]{\texttt{#1}}
\providecommand{\urlprefix}{URL }
\providecommand{\doi}[1]{https://doi.org/#1}

\bibitem{ahmed2024automatic}
Ahmed, T., Pai, K.S., Devanbu, P., Barr, E.: Automatic semantic augmentation of language model prompts (for code summarization). In: Proceedings of the IEEE/ACM 46th International Conference on Software Engineering. pp. 1--13 (2024)

\bibitem{howunsaferust20}
Astrauskas, V., Matheja, C., Poli, F., M\"{u}ller, P., Summers, A.J.: How do programmers use unsafe rust? Proc. ACM Program. Lang.  \textbf{4}(OOPSLA) (nov 2020). \doi{10.1145/3428204}, \url{https://doi.org/10.1145/3428204}

\bibitem{Bahmann2015ddg}
Bahmann, H., Reissmann, N., Jahre, M., Meyer, J.C.: Perfect reconstructability of control flow from demand dependence graphs. ACM Trans. Archit. Code Optim.  \textbf{11}(4) (jan 2015). \doi{10.1145/2693261}, \url{https://doi.org/10.1145/2693261}

\bibitem{bang2023trust}
Bang, I., Kayondo, M., Moon, H., Paek, Y.: {TRUST}: A compilation framework for in-process isolation to protect safe rust against untrusted code. In: 32nd USENIX Security Symposium (USENIX Security). Baltimore, MD, USA. (2023)

\bibitem{bui2021infercode}
Bui, N.D., Yu, Y., Jiang, L.: Infercode: Self-supervised learning of code representations by predicting subtrees. In: 2021 IEEE/ACM 43rd International Conference on Software Engineering (ICSE). pp. 1186--1197. IEEE (2021)

\bibitem{cppinitfiasco}
{C++ Reference}: Static initialization order fiasco, \url{https://en.cppreference.com/w/cpp/language/siof}

\bibitem{casey2024}
casey: Awesome rewrite it in rust (2024), \url{https://github.com/casey/awesome-rewrite-it-in-rust}, a curated list of replacements for existing software written in Rust

\bibitem{chen2021my}
Chen, Q., Xia, X., Hu, H., Lo, D., Li, S.: Why my code summarization model does not work: Code comment improvement with category prediction. ACM Transactions on Software Engineering and Methodology (TOSEM)  \textbf{30}(2),  1--29 (2021)

\bibitem{rust_uf}
Chenghao~Li, Y.W.: Demystifying compiler unstable feature usage and impacts in the rust ecosystem. In: Proceedings of the International Conference on Software Engineering (ICSE) (2024)

\bibitem{rust_clippy}
Chunmiao~Li, Y.Y.: Unleashing the power of clippy in real-world rust projects. In: Proceedings of the International Conference on Software Engineering (ICSE) (2024)

\bibitem{complexity_crate}
Crates.io: Complexity---calculate cognitive complexity of rust code (2024), \url{https://crates.io/crates/complexity}

\bibitem{rustassistance}
Deligiannis, P., Lal, A., Mehrotra, N., Rastogi, A.: Fixing rust compilation errors using llms. arXiv preprint arXiv:2308.05177  (2023), \url{https://arxiv.org/abs/2308.05177}

\bibitem{ibm_endianess}
Developer, I.: Introduction to big endian and little endian (2024), \url{https://developer.ibm.com/articles/au-endianc/#end}, accessed: 2024-05-30

\bibitem{arm_complex_macros}
developer.arm.com: Complex macros in c (2024), \url{https://developer.arm.com/documentation/101655/latest/Cx51-User-s-Guide/Preprocessor/Macros/Complex-Macros}

\bibitem{amd_pointer_aliasing}
Documentation, A.: Pointer aliasing (2024), \url{https://docs.amd.com/r/en-US/ug1079-ai-engine-kernel-coding/Pointer-Aliasing}, accessed: 2024-05-30

\bibitem{du2024mercury}
Du, M., Luu, A.T., Ji, B., Ng, S.K.: Mercury: An efficiency benchmark for llm code synthesis. arXiv preprint arXiv:2402.07844  (2024)

\bibitem{laertes23}
Emre, M., Boyland, P., Parekh, A., Schroeder, R., Dewey, K., Hardekopf, B.: Aliasing limits on translating c to safe rust. Proceedings of the ACM on Programming Languages  \textbf{7}(OOPSLA1),  551--579 (2023)

\bibitem{laertes21}
Emre, M., Schroeder, R., Dewey, K., Hardekopf, B.: Translating c to safer rust. Proceedings of the ACM on Programming Languages  \textbf{5}(OOPSLA),  1--29 (2021)

\bibitem{eniser2024towards}
Eniser, H.F., Zhang, H., David, C., Wang, M., Christakis, M., Paulsen, B., Dodds, J., Kroening, D.: Towards translating real-world code with {LLMs}: A study of translating to rust. arXiv preprint arXiv:2405.11514  (2024)

\bibitem{lev23}
Fiala, J., Itzhaky, S., M{\"u}ller, P., Polikarpova, N., Sergey, I.: Leveraging rust types for program synthesis. Proceedings of the ACM on Programming Languages  \textbf{7}(PLDI),  1414--1437 (2023)

\bibitem{rust_lifetime_2023}
Forum, R.U.: Lifetime issue: Slice of slices. \url{https://users.rust-lang.org/t/lifetime-issue-slice-of-slices/80848} (2023), accessed: 2024-06-08

\bibitem{gprof}
Graham, S.L., Kessler, P.B., McKusick, M.K.: Gprof: A call graph execution profiler. ACM Sigplan Notices  \textbf{17}(6),  120--126 (1982), \url{https://sourceware.org/binutils/docs/gprof/}

\bibitem{gu2023llm}
Gu, Q.: Llm-based code generation method for golang compiler testing. In: Proceedings of the 31st ACM Joint European Software Engineering Conference and Symposium on the Foundations of Software Engineering. pp. 2201--2203 (2023)

\bibitem{DBLP:journals/corr/abs-2309-08221}
Guo, Q., Cao, J., Xie, X., Liu, S., Li, X., Chen, B., Peng, X.: Exploring the potential of {ChatGPT} in automated code refinement: An empirical study. CoRR  \textbf{abs/2309.08221} (2023). \doi{10.48550/ARXIV.2309.08221}, \url{https://doi.org/10.48550/arXiv.2309.08221}

\bibitem{hong2024don}
Hong, J., Ryu, S.: Don’t write, but return: Replacing output parameters with algebraic data types in {C-to-Rust} translation. Proceedings of the ACM on Programming Languages  \textbf{8}(PLDI),  716--740 (2024)

\bibitem{hong2024tag}
Hong, J., Ryu, S.: To tag, or not to tag: Translating {C's} unions to {Rust's} tagged unions. In: Proceedings of the 39th IEEE/ACM International Conference on Automated Software Engineering. pp. 40--52 (2024)

\bibitem{hong2025type}
Hong, J., Ryu, S.: Type-migrating {C-to-Rust} translation using a large language model. Empirical Software Engineering  \textbf{30}(1), ~3 (2025)

\bibitem{c2rustfaq}
Immunant, I.: c2rust - c to rust translation and more (2024), \url{https://github.com/immunant/c2rust#faq}, accessed: 2024-06-03

\bibitem{infoq2023rusttrend}
InfoQ: Rust reviewed: the current trends and pitfalls of the ecosystem (2023), \url{https://www.infoq.com/articles/rust-ecosystem-review-2023/}, a review of the current trends and pitfalls in the Rust ecosystem

\bibitem{gpt4o}
Islam, R., Moushi, O.M.: Gpt-4o: The cutting-edge advancement in multimodal llm. Authorea Preprints  (2024)

\bibitem{rustbelt17}
Jung, R., Jourdan, J.H., Krebbers, R., Dreyer, D.: Rustbelt: Securing the foundations of the rust programming language. Proceedings of the ACM on Programming Languages  \textbf{2}(POPL),  1--34 (2017)

\bibitem{RustBookUnsafe}
Klabnik, S., Nichols, C.: The rust programming language - unsafe rust (2024), \url{https://doc.rust-lang.org/book/ch19-01-unsafe-rust.html#accessing-or-modifying-a-mutable-static-variable}, accessed: 2024-05-30

\bibitem{crate-goto}
Kotlyarov, D.: Crate goto (2024), \url{https://docs.rs/goto/latest/goto/}

\bibitem{rust_lang_unsafe_rust}
rust lang.org: Unsafe rust (2023), \url{https://doc.rust-lang.org/book/ch19-01-unsafe-rust.html}, accessed: 2023-11-18

\bibitem{rust_forum_lazy_static}
Lazytiger, et~al.: What is the proper way to use lazy\_static variables safely and briefly (2023), \url{https://users.rust-lang.org/t/what-is-the-proper-way-to-use-lazy-static-variables-safely-and-briefly/94739}, accessed: 2024-05-27

\bibitem{learncpp_switch_fallthrough_scoping}
LearnCpp.com: Switch fallthrough and scoping. \url{https://www.learncpp.com/cpp-tutorial/switch-fallthrough-and-scoping/} (2024), accessed: 2024-06-08

\bibitem{crusts}
Ling, M., Yu, Y., Wu, H., Wang, Y., Cordy, J.R., Hassan, A.E.: In rust we trust: a transpiler from unsafe c to safer rust. In: Proceedings of the ACM/IEEE 44th International Conference on Software Engineering: Companion Proceedings. pp. 354--355 (2022)

\bibitem{liu2024your}
Liu, J., Xia, C.S., Wang, Y., Zhang, L.: Is your code generated by chatgpt really correct? rigorous evaluation of large language models for code generation. Advances in Neural Information Processing Systems  \textbf{36} (2024)

\bibitem{rusthornbelt22}
Matsushita, Y., Denis, X., Jourdan, J.H., Dreyer, D.: Rusthornbelt: a semantic foundation for functional verification of rust programs with unsafe code. In: Proceedings of the 43rd ACM SIGPLAN International Conference on Programming Language Design and Implementation. pp. 841--856 (2022)

\bibitem{c2rust_challenges}
Mertens, E., Langley, T.: {C2Rust}: Migrating {C} to safe {Rust} (2018), \url{https://galois.com/blog/2018/08/c2rust/}, accessed: 2024-05-24

\bibitem{rust_unsafe}
Mohan~Cui, S.S.: Is unsafe an achilles' heel? a comprehensive study of safety requirements in unsafe rust programming. In: Proceedings of the International Conference on Software Engineering (ICSE) (2024)

\bibitem{rosettacode}
Mol, M.: Rosetta code, \url{https://rosettacode.org}, accessed: 2024-05-29

\bibitem{nam2024using}
Nam, D., Macvean, A., Hellendoorn, V., Vasilescu, B., Myers, B.: Using an llm to help with code understanding. In: Proceedings of the IEEE/ACM 46th International Conference on Software Engineering. pp. 1--13 (2024)

\bibitem{openai2024gpt4}
OpenAI: {GPT-4} technical report. Tech. rep., OpenAI (2024)

\bibitem{pan2024lost}
Pan, R., Ibrahimzada, A.R., Krishna, R., Sankar, D., Wassi, L.P., Merler, M., Sobolev, B., Pavuluri, R., Sinha, S., Jabbarvand, R.: Lost in translation: A study of bugs introduced by large language models while translating code. In: Proceedings of the IEEE/ACM 46th International Conference on Software Engineering. pp. 1--13 (2024)

\bibitem{peterson1973capabilities}
Peterson, W.W., Kasami, T., Tokura, N.: On the capabilities of while, repeat, and exit statements. Communications of the ACM  \textbf{16}(8),  503--512 (1973)

\bibitem{cflowgnu}
Poznyakoff, S.: Gnu cflow (2005), \url{https://www.gnu.org/software/cflow/}

\bibitem{memorysafetyrust2020}
Qin, B., Chen, Y., Yu, Z., Song, L., Zhang, Y.: Understanding memory and thread safety practices and issues in real-world rust programs. In: Proceedings of the 41st ACM SIGPLAN Conference on Programming Language Design and Implementation. p. 763–779. PLDI 2020, Association for Computing Machinery, New York, NY, USA (2020). \doi{10.1145/3385412.3386036}, \url{https://doi.org/10.1145/3385412.3386036}

\bibitem{Ramsey2022relooper}
Ramsey, N.: Beyond relooper: recursive translation of unstructured control flow to structured control flow (functional pearl). Proc. ACM Program. Lang.  \textbf{6}(ICFP) (aug 2022). \doi{10.1145/3547621}, \url{https://doi.org/10.1145/3547621}

\bibitem{commonbugFixRust}
Robati~Shirzad, M., Lam, P.: A study of common bug fix patterns in rust. Empirical Software Engineering  \textbf{29}(2), ~44 (2024)

\bibitem{bzip2}
Seward, J.: bzip2 (2023), \url{https://sourceware.org/bzip2/}, accessed: 2024-06-04

\bibitem{SyRust2021}
Takashima, Y., Martins, R., Jia, L., P\u{a}s\u{a}reanu, C.S.: Syrust: automatic testing of rust libraries with semantic-aware program synthesis. In: Proceedings of the 42nd ACM SIGPLAN International Conference on Programming Language Design and Implementation. p. 899–913. PLDI 2021, Association for Computing Machinery, New York, NY, USA (2021). \doi{10.1145/3453483.3454084}, \url{https://doi.org/10.1145/3453483.3454084}

\bibitem{TaKO8Ki2024}
TaKO8Ki: Awesome alternatives in rust (2024), \url{https://github.com/TaKO8Ki/awesome-alternatives-in-rust}, a curated list of replacements for existing software written in Rust

\bibitem{tarjan1972depth}
Tarjan, R.: Depth-first search and linear graph algorithms. SIAM journal on computing  \textbf{1}(2),  146--160 (1972)

\bibitem{rust_lancet}
Wenzhang~Yang, L.S.: Rust-lancet: Automated ownership-rule-violation fixing with behavior preservation. In: Proceedings of the International Conference on Software Engineering (ICSE) (2024)

\bibitem{white2023chatgpt}
White, J., Other, A.: {ChatGPT} prompt patterns for improving code quality, refactoring, requirements elicitation, and software design. arXiv preprint arXiv:2303.07839  (2023)

\bibitem{cyclomatic_complexity}
Wikipedia: Cyclomatic complexity (2024), \url{https://en.wikipedia.org/wiki/Cyclomatic_complexity}

\bibitem{redox}
Wikipedia: Redox (operating system) (2024), \url{https://en.wikipedia.org/wiki/Redox_(operating_system)}

\bibitem{rust4linux}
Wikipedia: Rust for linux (2024), \url{https://en.wikipedia.org/wiki/Rust_for_Linux}

\bibitem{wikipedia_type_punning}
Wikipedia: Type punning (2024), \url{https://en.wikipedia.org/wiki/Type_punning#C_and_C++}

\bibitem{rust_security}
Xiaoye~Zheng, Z.W.: A closer look at the security risks in the rust ecosystem. ACM Transactions on Software Engineering and Methodology (TOSEM)  (2024)

\bibitem{xu2022systematic}
Xu, F.F., Alon, U., Neubig, G., Hellendoorn, V.J.: A systematic evaluation of large language models of code. In: Proceedings of the 6th ACM SIGPLAN International Symposium on Machine Programming. pp. 1--10 (2022)

\bibitem{yang2024exploring}
Yang, Z., Liu, F., Yu, Z., Keung, J.W., Li, J., Liu, S., Hong, Y., Ma, X., Jin, Z., Li, G.: Exploring and unleashing the power of large language models in automated code translation. arXiv preprint arXiv:2404.14646  (2024)

\bibitem{rust_klee}
Ying~Zhang, P.L.: Broadly enabling klee to effortlessly find unrecoverable errors in rust. In: Proceedings of the International Conference on Software Engineering (ICSE) (2024)

\bibitem{rust_oss}
Yuxia~Zhang, M.Q.: How are paid and volunteer open source developers different? a study of the rust project. In: Proceedings of the International Conference on Software Engineering (ICSE) (2024)

\bibitem{zhang2024scalable}
Zhang, H., David, C., Wang, M., Paulsen, B., Kroening, D.: Scalable, validated code translation of entire projects using large language models. arXiv preprint arXiv:2412.08035  (2024)

\bibitem{crown}
Zhang, H., David, C., Yu, Y., Wang, M.: Ownership guided {C to Rust} translation. arXiv preprint arXiv:2303.10515  (2023)

\bibitem{rust_rpg}
Zhiwu~Xu, B.W.: {RPG}: Rust library fuzzing with pool-based fuzz target generation and generic support. In: Proceedings of the International Conference on Software Engineering (ICSE) (2024)

\end{thebibliography}

\end{document}